# The role of Al$_2$O$_3$ interlayer in the synthesis of ZnS/Al$_2$O$_3$/MoS$_2$ core-shell nanowires

*Edgars Butanovs[1,3], Alexei Kuzmin[1], Aleksejs Zolotarjovs[1], Sergei Vlassov[2], Boris Polyakov[1*]*

[1]Institute of Solid State Physics, University of Latvia, Kengaraga street 8, LV-1063 Riga, Latvia

[2]Institute of Physics, University of Tartu, W. Ostwaldi Str. 1, 50412, Tartu, Estonia

[3]Institute of Technology, University of Tartu, Nooruse 1, 50411 Tartu, Estonia

*e-mail: boris.polyakov@cfi.lu.lv

**Abstract**

During the synthesis of heterostructured nanomaterials, unwanted structural and morphological changes in nanostructures may occur, especially when multiple sequential growth steps are involved. In this study, we describe a synthesis strategy of heterostructured ZnS/Al$_2$O$_3$/MoS$_2$ core-shell nanowires (NWs), and explore the role of the Al$_2$O$_3$ interlayer during synthesis. Core-shell NWs were produced via a four-step route: 1) synthesis of ZnO NWs on a silicon wafer, 2) deposition of thin Al$_2$O$_3$ layer by ALD, 3) magnetron deposition of MoO$_3$ layer, and 4) annealing of the sample in the sulphur atmosphere. During sulphurization, ZnO is converted into ZnS, and MoO$_3$ into MoS$_2$, while the Al$_2$O$_3$ interlayer preserves the smooth surface of an NW required for the growth of a continuous MoS$_2$ shell. The resulting ZnS/Al$_2$O$_3$/MoS$_2$ core-shell NWs were characterized by transmission electron microscopy, X-ray diffraction and photoelectron spectroscopy, Raman spectroscopy, and optical photoluminescence spectroscopy. A reported strategy can be used for the synthesis of other core-shell NWs with a transition metal dichalcogenides (TMDs) shell to protect the NW core material that may otherwise be altered or damaged by the reactive chalcogenides at high temperatures.

## 1. Introduction

In recent years, two-dimensional (2D) transition metal dichalcogenides (TMDs) have become a topic of interest within the materials-research scientific community [1]. TMDs with the general formula $MX_2$ (e.g., M = W, Mo, etc., X = S, Se, Te) have an indirect bandgap in bulk form but transform to direct energy bandgap semiconductors when scaled down to a few atomic layers [2,3]. Molybdenum disulfide ($MoS_2$) is one of the most studied and well-known 2D TMDs materials [4], and can be synthesized in the form of powder, few-layer or monolayer 2D nanosheets, nanoflakes, and nanotubes [5]. $MoS_2$ is an n-type semiconductor; however, the origin of this unintentional doping is still not clear [6]. $MoS_2$ has the indirect $E_{ig} = 1.2$ eV and direct $E_{dg} = 1.8$ eV bandgap in bulk and monolayer forms, respectively, and is non-toxic and thermally and chemically stable [7]. Some of the physical properties of $MoS_2$, including a direct bandgap in low-dimensional structures, strong interaction of light with matter, and good carrier mobility, have aroused interest in this material in the field of optoelectronics and photodetection [8]. On the other hand, $MoS_2$ exhibits excellent catalytic properties and good potential for photocatalytic production of solar fuels like hydrogen gas [9]. A number of studies are dedicated to the synthesis and characterization of hybrid nanostructures that incorporate layered $MoS_2$. [10–15] In particular, $MoS_2$ is often combined with ZnO [10,14–19] and ZnS [10,20–22], governed by the attractive properties of these materials.

Bulk zinc oxide (ZnO) is a direct bandgap (3.4 eV) n-type semiconductor [23] and has two main phases: hexagonal wurtzite (P63mc) and cubic (Fm3m) zincblende. Nanostructured ZnO has been of interest to the scientific community for decades, and is still very popular due to the simplicity of its synthesis (no need to use expensive raw materials or complex equipment) and numerous morphological shapes and nanostructures (NWs, nanobelts, nanotubes, etc.), which make it very attractive as a nano-template material [24].

Zinc sulfide (ZnS) is a direct bandgap semiconductor with a bandgap of 3.6 eV for the most stable zincblende phase, and 3.9 eV for the high-temperature wurtzite phase [25]. ZnS is a low-cost, environmentally friendly compound with attractive mechanical properties, such as good fracture strength and hardness. ZnS is transparent in a wide wavelength range and is widely used for electroluminescent applications and nanophosphors [26]. ZnS NWs have shown promise in electronics, optoelectronics, and sensor applications [27].

Core-shell nanowire-based architecture is highly promising for different solar energy harvesting applications [28,29]. ZnO/$MoS_2$ hybrid heterostructures are promising for photocatalytic reactions, including the hydrogen evolution reaction [10,14–19]. These [10,20–22] nanostructures are also shown to possess favorable optical properties, including enhanced photoelectric performance [30], making them potentially suitable for imaging and sensing applications. In addition, we recently demonstrated that it is possible to combine all three materials and their useful properties into a single ZnO/ZnS/$MoS_2$ core-shell heterostructure [10]; however, we found that during high-temperature synthesis of ZnS/$MoS_2$ core-shell NWs, ZnO is partially or completely converted into ZnS, which causes severe morphological changes in the NW core [10].

$Al_2O_3$ is an environmentally-friendly and technologically important material with high hardness and chemical stability [31], and can exist in many crystallographic modifications possessing different properties. The reported experimental band gap value is 8.8 eV for α-$Al_2O_3$, 7.0−8.7 eV for γ-$Al_2O_3$, and 5.1−7.1 eV for amorphous $Al_2O_3$ [32]. A thin layer of amorphous $Al_2O_3$ can be deposited with high uniformity and thickness accuracy using the atomic layer deposition (ALD) technique at low temperatures [33]. Furthermore, it has been shown that a thin $Al_2O_3$ layer can be used for the chemical and mechanical protection of metallic nanowires (NWs) [34].

In this study, we report on the synthesis of ZnS/$Al_2O_3$/$MoS_2$ core-shell NWs and demonstrate the protective role of the $Al_2O_3$ interlayer during the chemical conversion of ZnO and $MoO_3$ into ZnS and $MoS_2$, respectively, during annealing in the sulphur atmosphere. The choice of the $Al_2O_3$ interlayer can be explained by three main factors: 1) $Al_2O_3$ material has high chemical stability in the sulphur environment at elevated temperatures; 2) the ALD deposition process of $Al_2O_3$ is a well-established technological process, which allows controlling shell thickness with sub-nanometer precision; and 3) $MoS_2$ and other TMDs materials were successfully grown on both crystalline and amorphous $Al_2O_3$ substrates. ZnO NWs are convenient 1D templates for core-shell synthesis of nanomaterials. A combination of ZnO NWs with an $Al_2O_3$ interlayer may be useful for many other TMDs materials synthesis, where the sacrificial precursor layer is selenized or sulphurized at elevated temperatures.

2. Experimental details.

$ZnS/Al_2O_3/MoS_2$ core-shell NWs were produced via a four-step route: 1) synthesis of ZnO NWs on a silicon wafer; 2) deposition of thin $Al_2O_3$ layer by ALD; 3) magnetron deposition of the $MoO_3$ layer; and 4) annealing of the sample in the sulphur atmosphere.

ZnO NWs were grown by a chemical vapor transport method using spherical Au nanoparticles as a catalyst (50 nm in diameter, water suspension, *BBI International*) [7]. A 1:4 mixture of ZnO and graphite powders was heated to 900 °C in a quartz tube for 30 minutes in a stream of the carrier $N_2$ gas. NWs were synthesized on top of the oxidized silicon wafers $Si/SiO_2$ (Si (100) wafer, 50 nm of thermal oxide, *Semiconductor Wafer, Inc.*). $Al_2O_3$ films were grown in the ALD reactor Savannah S100. The films were grown at 150 °C by applying 66 cycles of Trimethylaluminum (TMA) and $H_2O$ as precursors, and inert gas ($N_2$) flow. Next, $ZnO/Al_2O_3$ NW samples were coated with a layer of amorphous a-$MoO_3$, having a 30 nm thickness on a flat substrate, and using reactive dc magnetron sputtering of metallic tungsten target in a mixed $Ar/O_2$ atmosphere. Finally, the samples were annealed in a quartz tube in a sulphur atmosphere using $N_2$ as a carrier gas for 30 minutes at 750 °C to convert the molybdenum precursor ($MoO_3$) into $MoS_2$.

2D microcrystals of $MoS_2$ used as the reference sample for micro-Raman measurements were grown for 20 minutes on a $Si/SiO_2$ substrate in a quartz tube using 1 mg $MoO_3$ (99.5%,

*Sigma Aldrich*) and 0.25 g sulfur (99.98%, *Sigma Aldrich*) powders as precursor materials and $N_2$ as a carrier gas. Silicon substrate was placed downstream of precursors in the high-temperature zone (700 °C).

ZnS NWs were synthesized as reference material for XRD and PL measurements. NWs were synthesized on top of the oxidized silicon wafers $Si/SiO_2$ using the same Au nanoparticles as a catalyst. ZnS powder 0.4 g (> 97%, *Sigma Aldrich*) was thermally sublimated in a quartz tube at a temperature of 950 °C for 30 minutes, followed by natural cooling. The ZnS vapor was carried downstream by $N_2$ gas to the substrate to grow ZnS NWs.

The phase composition of the samples was studied with X-ray diffraction (XRD) using a Rigaku MiniFlex 600 X-ray powder diffractometer, Bragg-Brentano θ-2θ geometry, and a 600 W Cu anode (Cu Kα radiation, λ = 1.5406 Å) X-ray tube. The inner crystalline structure of core-shell NWs transferred on the Lacey Cu TEM grid (*Agar Scientific*) was revealed using TEM (Tecnai GF20, FEI) operated at the accelerating voltage of 200 kV.

X-ray photoelectron spectroscopy (XPS) measurements were performed using an ESCALAB Xi (ThermoFisher) X-ray photoelectron spectrometer. The spectra were calibrated relative to the adventitious C 1s peak at 284.8 eV.

Micro-Raman spectra were measured using a TriVista 777 confocal Raman system (Princeton Instruments/Acton, 750 mm focal length, 600 lines/mm grating) equipped with an upright Olympus microscope with an Olympus UIS2 LMPlanFLN 20x/0.40 objective. The Raman spectra were excited by a continuous-wave, single-frequency, diode-pumped laser Cobolt Samba 150 (λ = 532 nm) and recorded with an Andor iDus DV420A-OE CCD camera. The laser power on the sample was controlled by a set of neutral density filters and was set to about 2.1 mW.

Imaging experiments were performed at room temperature using a confocal microscope with spectrometer Nanofinder-S (SOLAR TII) [35,36]. The confocal reflection and spectral images were recorded in a back-scattering geometry with a Nikon CF Plan Apo 100× (NA=0.95) optical objective in a raster-scanning mode. A diode-pumped solid-state (DPSS) Nd: YAG laser (532 nm, max continuous wave (cw) power Pex=150 mW) was used as the photoluminescence (PL) excitation source. The PL signal was dispersed by 150 grooves/mm diffraction grating mounted in the 520 mm focal length monochromator, and two Hamamatsu R928 photomultiplier tubes were used as detectors. To avoid sample damage, the laser power at the sample was controlled by a variable neutral filter.

Room-temperature photoluminescence (PL, Hamamatsu R92P PMT) spectra with a 266 nm excitation wavelength (fourth harmonic of CryLas Nd:YAG laser, 0.3 mJ power, 1 ns pulse duration, 5 kHz repetition rate) were measured to investigate the as-prepared nanostructure optical properties.

The thickness of MoS2 2D microcrystals was measured with an atomic force microscope (AFM, CP-II, Veeco) in the tapping mode using PPP-NCHR probes with a force constant of 42 N/m and tip radius of curvature < 10 nm (Nanosensors).

3. Results and discussion.

The phase composition of NWs grown on oxidized Si substrates and annealed in the air at 750 °C ($ZnO/Al_2O_3$) after sulphurisation ($ZnS/MoS_2$ and $ZnS/Al_2O_3/MoS_2$) was studied using the XRD method (Fig. 1). According to XRD data, no ZnO phase was present in NWs after sulphurisation, meaning ZnO was completely converted to the ZnS phase. Such behavior was expected since sulphur reacts with ZnO (ICDD 36-1451) at temperatures above 400 °C, resulting in the formation of the ZnS phase (ICDD 36-1450) [37]. No Bragg peaks of the $MoO_3$ phase were found in NWs covered with a molybdenum oxide shell, but the (002) $MoS_2$ peak (ICDD 37-1492) was detected, indicating the full conversion of $MoO_3$ into $MoS_2$ during sulphurisation. Also, no peaks of $Al_2O_3$ were observed, likely due to the small thickness of the $Al_2O_3$ shell. Several other Bragg peaks present on XRD patterns were attributed to the Si(100) substrate (forbidden Si(200) reflection at 2θ ≈ 33°), cristobalite SiO$_2$ (ICDD 36-1451), and the gold nanoparticles used for VLS growth (ICDD 04-0784).

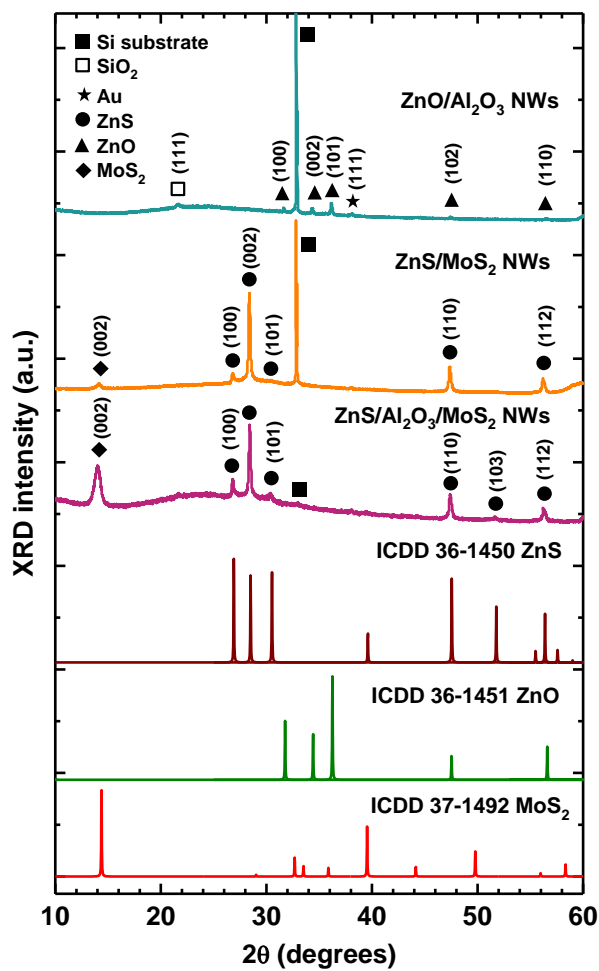

*Figure 1*. XRD patterns of NWs annealed in the air (ZnO/Al$_2$O$_3$) and after sulphurisation (ZnS/MoS$_2$ and ZnS/Al$_2$O$_3$/MoS$_2$).

In addition to XRD measurements, XPS analysis was performed on the as-prepared NWs to verify the elemental composition and study the chemical states of the component ions in the nanostructures. XPS spectra for the ZnS/MoS$_2$ and ZnS/Al$_2$O$_3$/MoS$_2$ NWs are shown in Fig. 2., while the spectra of the reference samples (ZnO, ZnO-ZnS, ZnS, ZnO-Al$_2$O$_3$, and ZnO-Al$_2$O$_3$-MoO$_3$ NWs), which were used to determine peak relative shifts and perform detailed chemical-state analysis of the various compounds, are provided in the Supplementary Information in Fig.S1.

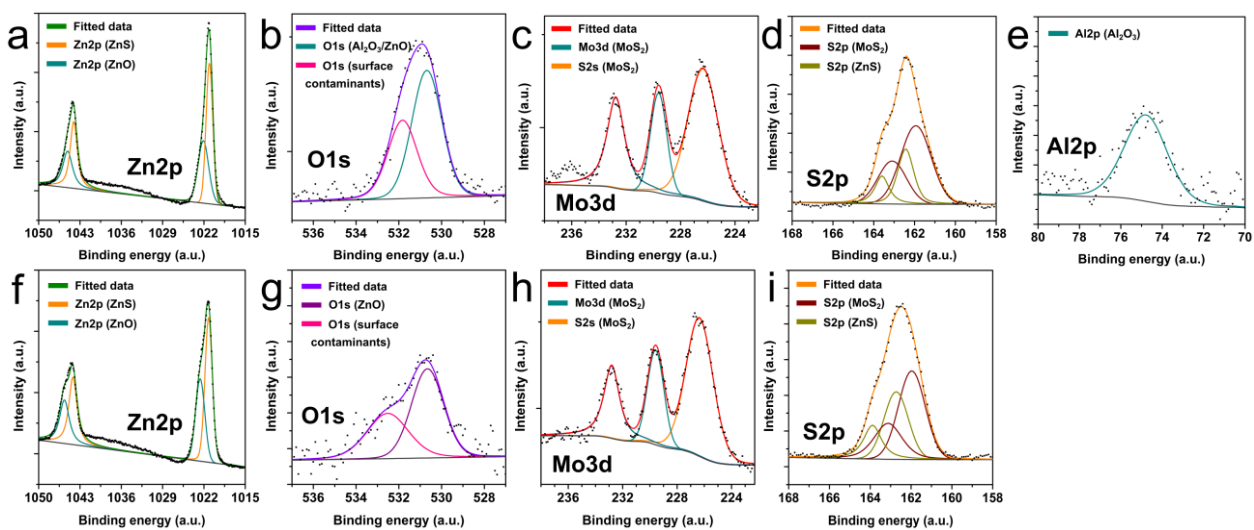

*Figure 2*. High-resolution XPS spectra and peak fits for elements in the sulphurized NWs (a-e) ZnS/Al$_2$O$_3$/MoS$_2$ and (f-i) ZnS/MoS$_2$ NWs.

The composition analysis confirmed the presence of the expected elements in the samples and did not show any impurities. The high-resolution scans of each element's characteristic peaks provided information on the chemical states. Zinc Zn 2p scans (see Fig. 2(a,f)) gave qualitatively similar results for both samples: a pair of Zn 2p$_{3/2}$ (at 1021.2 eV binding energy) and 2p$_{1/2}$ peaks with spin-orbit splitting $\Delta_{3/2-1/2}$=23 eV was attributed to ZnS compound, and a pair of significantly weaker peaks shifted slightly to a higher energy (Zn 2p$_{3/2}$ at 1022.1 eV) was attributed to ZnO, thus showing a remaining amount of the compound in the NWs, which XRD measurements did not detect. The shift between the Zn compounds was also observed in the ZnO, ZnO-ZnS, and ZnS reference samples (see Fig.S1(a,c,e)) as was previously reported in the literature [38,39]. Oxygen O 1s scans Fig.2(b,g) show the typical metal oxide peak centered at around 530.4 eV with a broader shoulder at 532-533 eV attributed to organic contaminants on the surface (C-O and C=O bonds). Both Al$_2$O$_3$ and ZnO compounds contribute to the peak in the ZnS/Al$_2$O$_3$/MoS$_2$ sample, while weak ZnO contribution was observed in the ZnS/MoS$_2$ sample. Molybdenum Mo 3d scans (Fig.2(c,h)) revealed three peaks: two Mo 3d 5/2 (at 229.6 eV) and 3/2 peaks with 3.2 eV spin-orbit splitting assigned to the MoS$_2$ compound, as well as an overlapping S 2s peak at 226.2 eV. An Mo 3d$_{5/2}$ peak at 233 eV (see Fig.S1(k)) for MoO$_3$ was not observed, thus confirming that the precursor MoO$_3$ coating was fully converted. Sulphur S 2p scans in Fig.2(d,i) show a broad peak around 162.3 eV, which consists of contributions from pairs of S 2p 3/2 and 1/2 ($\Delta_{3/2-1/2}$=1.16 eV) for both $MoS_2$ and ZnS compounds. The Zn peaks for ZnS are shifted slightly to a higher energy compared to the peaks for $MoS_2$, as has been reported before [40]. Finally, Al 2p peak (see Fig.2(e)) at 74.8 eV is typically associated with the $Al_2O_3$ compound. The XPS measurements clearly show that the introduction of the $Al_2O_3$ interlayer does not affect the chemistry of the desired compounds in the heterostructures, as the XPS results for the samples with and without $Al_2O_3$ have qualitatively similar results.

Raman spectroscopy was used to fingerprint the presence of the $MoS_2$ phase. The obtained Raman spectra of the core-shell NWs (Fig. 3) confirm the formation of the $MoS_2$ layer around the ZnS NW core.

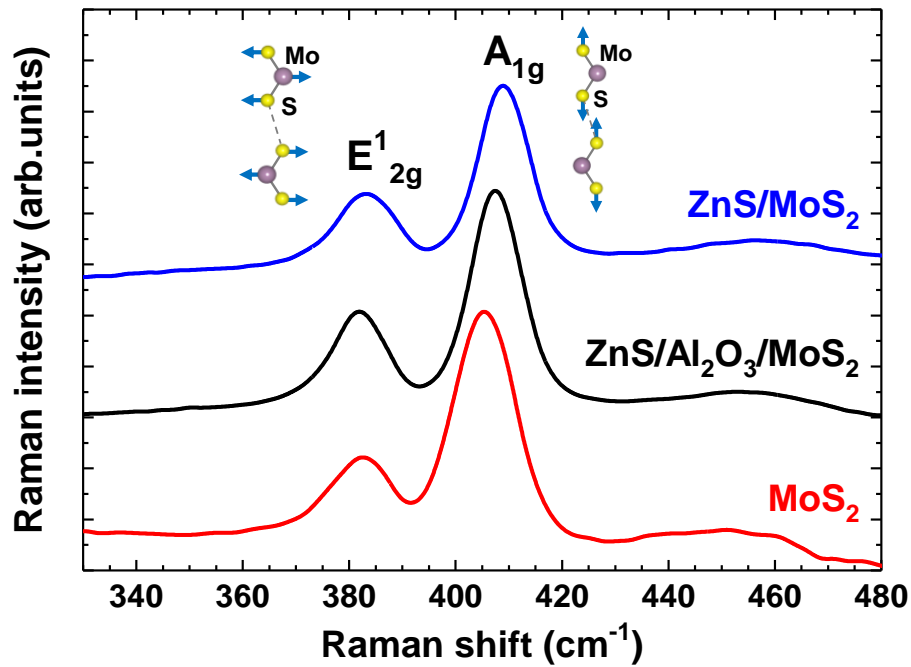

*Figure 3. Typical Raman spectra of $ZnS/MoS_2$ and $ZnS/Al_2O_3/MoS_2$ NWs and pure $MoS_2$. Two first-order Raman active modes (in-plane $E^1_{2g}$ and out-of-plane $A_{1g}$) are indicated.*

Two Raman-active zone center phonon modes, the in-plane $E^1_{2g}$ mode at 384 cm$^{-1}$ and the out-of-plane $A_{1g}$ mode at 407 cm$^{-1}$, are well resolved [41–44]. Raman bands due to ZnO [45], ZnS [46], and $Al_2O_3$ [47] phases were not observed. Note that the Raman spectrum of the $ZnS/MoS_2$ sample has a strong background of increasing towards a large Raman shift, i.e., long wavelength (not visible in the range shown in Fig. 3). The origin of this contribution is the photoluminescence from the direct excitonic transitions in a few-layer $MoS_2$ [2]. The absence of photoluminescence in the samples with the $Al_2O_3$ interlayer suggests that the presence of $Al_2O_3$ stimulates the growth of $MoS_2$ multilayers.

TEM microscopy was used to gain a deeper insight into the inner structure of core-shell NWs (Fig.4). As was reported in our previous work, sulfurization of $ZnO/MoO_3$ NWs at 700° C and conversion of ZnO into ZnS cause morphological changes [10]. With the increase in temperature up to 750°C, morphological changes in core-shell NW structures become more expressed (Fig. 4a-c). According to XRD data, after sulfurization, $ZnO/MoO_3$ NW converts into ZnS/MoS2, and MoS2 microcrystals can be distinguished at high magnification (Fig. 4c). As previously mentioned, a thin layer of amorphous $Al_2O_3$ was deposited by ALD as a protective shell around ZnO NWs. TEM images of $ZnO/Al_2O_3$ core-shell NW annealed in air at 750°C are shown in Fig. 4d-f at different magnifications. Complete crystallization of the $Al_2O_3$ shell occurs during NW annealing (Fig. 4c). According to the analysis of the SAED data, the $Al_2O_3$ layer grows in conformity to the m-plane of ZnO (Supplementary materials, Fig. S2). Due to absence of $Al_2O_3$ contribution in the XRD patterns, it is difficult to identify the crystalline phase of the $Al_2O_3$ shell (Supplementary materials, Fig. S3-4). According to the literature, ALD-deposited amorphous $Al_2O_3$ films can be crystallized in the alpha or gamma phase [48]. The possibility of epitaxial growth of m-plane ZnO on a (10-10) plane of $\alpha$-$Al_2O_3$ was reported by Ku et al. [49]. A sulfurized $ZnS/Al_2O_3/MoS_2$ core-shell NW is shown in Fig. 4g-i. The layers of $MoS_2$ grown from $MoO_3$ upon sulphurisation on the NW surface are distinguishable as parallel black lines. Typically, the thickness of the coating varies between 5 and 10 monolayers, with interlayer distance measured around 6.25 Å, which corresponds with the 6.2–6.3 Å interlayer distance in $MoS_2$ nanostructures [16,50]. We conclude that the $Al_2O_3$ interlayer was able to preserve the straight shape of the ZnO NW and facilitate the growth of $MoS_2$ parallel to the NW surface.

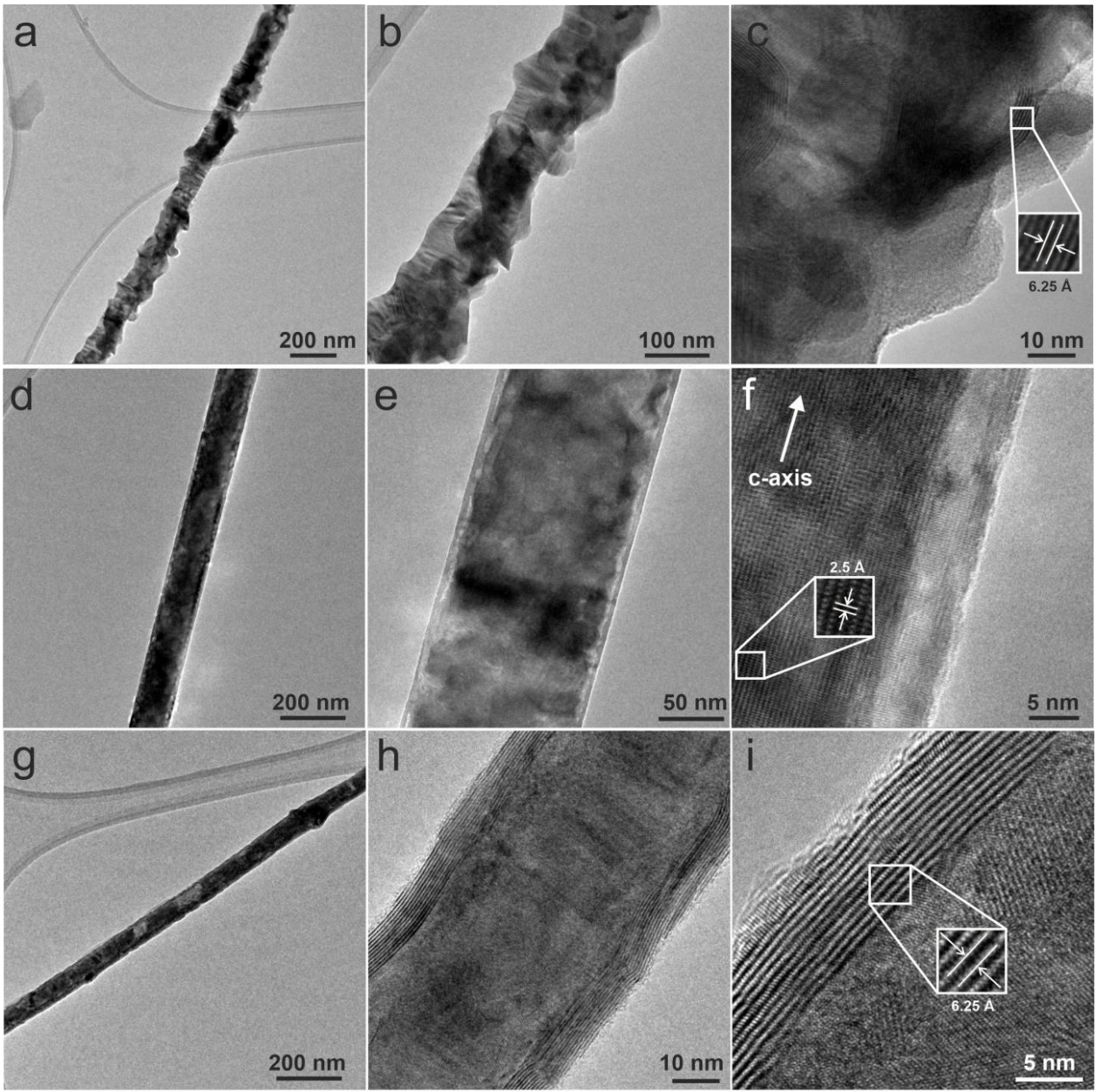

*Figure 4*. TEM images of ZnS/MoS$_2$ (a-c), ZnO/Al$_2$O$_3$ (d-f), and ZnS/Al$_2$O$_3$/MoS$_2$ (g-i) NWs annealed at 750°C.

It is an open question whether amorphous or crystalline Al$_2$O$_3$ is needed for the growth of qualitative MoS$_2$. According to Peters et al., MoS$_2$ 2D crystals can be grown on c-plane, r-plane, a-plane, and m- planes of α-Al$_2$O$_3$ substrates, but aligned growth occurs only on the r- and m-plane sapphire [51]. In the work of Ma et al., aligned rectangle-shaped MoS$_2$ 2D crystals were grown on the a-plane of sapphire [52], and high-quality MoS$_2$ 2D crystals were grown by CVD on an amorphous Al$_2$O$_3$ substrate by Bergeron et al. [53]. The crystallization temperature of α-

$MoO_3$ is about 340 °C [54], while the crystallization of amorphous $Al_2O_3$ was reported at temperatures above 650°C [55]. It is known that sulphuration of $MoO_3$ and conversion to $MoS_2$ begins at 500°C [10], meaning the $MoS_2$ layer likely starts to grow on the surface of the amorphous $Al_2O_3$ surface, and the main role of the $Al_2O_3$ interlayer is preserving the NW's straight shape.

Photoluminescence (PL) spectra of as-synthesized ZnS NWs, $ZnS/MoS_2$, and $ZnS/Al_2O_3/MoS_2$ NWs samples ($ZnO/MoO_3$ and $ZnO/Al_2O_3/MoO_3$ sulphurated at 750°C) measured at room temperature are shown in Fig. 5. The PL spectrum of ZnS NWs has a defect-related band at ~523 nm (2.37 eV) in good agreement with the literature [56,57]. For example, Chen et. al [57] attributes the PL band centered at 525 nm in the ZnS nanostructures to the Zn vacancies. Similarly, Wang et al. [56] attributed the peak at 2.40 eV (517 nm) to the presence of intrinsic vacancies such as the "S" vacancy acting as a donor center and the "Zn" vacancy acting as an acceptor. Up to 4 bands were observed in the experimental PL spectra of core-shell NWs and are described by Gaussian functions (shown as dotted lines in Fig. 5a). The PL spectrum of the $ZnS/Al_2O_3/MoS_2$ NWs contains both defect-related ZnS luminescence at ~520 nm (2.38 eV) and two luminescence bands related to direct excitonic transitions in $MoS_2$ at ~654 nm (1.90 eV) and ~707 nm (1.75 eV) [2]. An additional band at 580 nm (2.14 eV) can be seen in the PL spectrum of $ZnS/MoS_2$ NWs, which also appears as a "shoulder" in the $ZnS/Al_2O_3/MoS_2$ spectrum, the origin of which cannot be definitively identified. This band is related to sulfurized $ZnO/Al_2O_3$ NWs: a PL band in close wavelength spectral region ~616 nm (2.01 eV) can be seen for the $ZnO/Al_2O_3$ NWs sample sulfurized at 750° C (Fig. S6, Supplementary materials). On the other hand, ZnO crystals and ZnO nanomaterials have defect-related PL in the region of 2.2 eV [58]. It is possible that some amount of ZnO nanowires was not sulfurized completely to ZnS and gives a contribution to the PL spectra (note that XRD patterns did not contain any ZnO signal after sulfurization). In the PL spectrum of the $ZnS/MoS_2$ NWs, the band due to defect-related ZnS luminescence is reduced significantly, while $MoS_2$ related luminescence is similar to that in ZnS/Al$_2$O$_3$/MoS$_2$ NWs. Splendiani et al. observed the photoluminescence peaks for the MoS$_2$ monolayer at 1.98 eV (626 nm) and 1.83 eV (678 nm), corresponding to the direct excitonic transitions with the energy split from valence band spin-orbital coupling. Note that this photoluminescence is observed only for monolayers or a few layers of MoS$_2$, and is absent in the indirect bandgap bulk MoS$_2$ sample. Confocal microscope images of MoS$_2$ 2D crystals grown on the Si/SiO$_2$ wafer are shown in Fig. 5b,c. The image in Fig. 5b was taken in the reflection mode for 532 nm, and the intensity of the gray color is proportional to the sample reflectivity, therefore, thicker regions of MoS$_2$ appear darker. The light color of the regions surrounding MoS$_2$ 2D crystals is due to the reflectivity of the Si substrate, whereas the bright regions at the center of the two MoS$_2$ 2D crystals are due to the high reflectivity of thick metallic-like MoS$_2$. The image in Fig. 5c is a photoluminescence map that was taken simultaneously with the reflectivity image in Fig. 5b. The PL was excited by a 532 nm laser line and measured at 685 nm, and the PL map correlates well with the reflectivity image. According to AFM measurements (Supplementary, Fig. S5), 2D MoS$_2$ crystals are thicker in the center (up to 250 nm) and thinner at the periphery (2-3 nm). There is no PL signal in the central region of 2D MoS$_2$ crystals (Fig. 5c) due to a transition from direct to indirect bandgap at higher thickness.

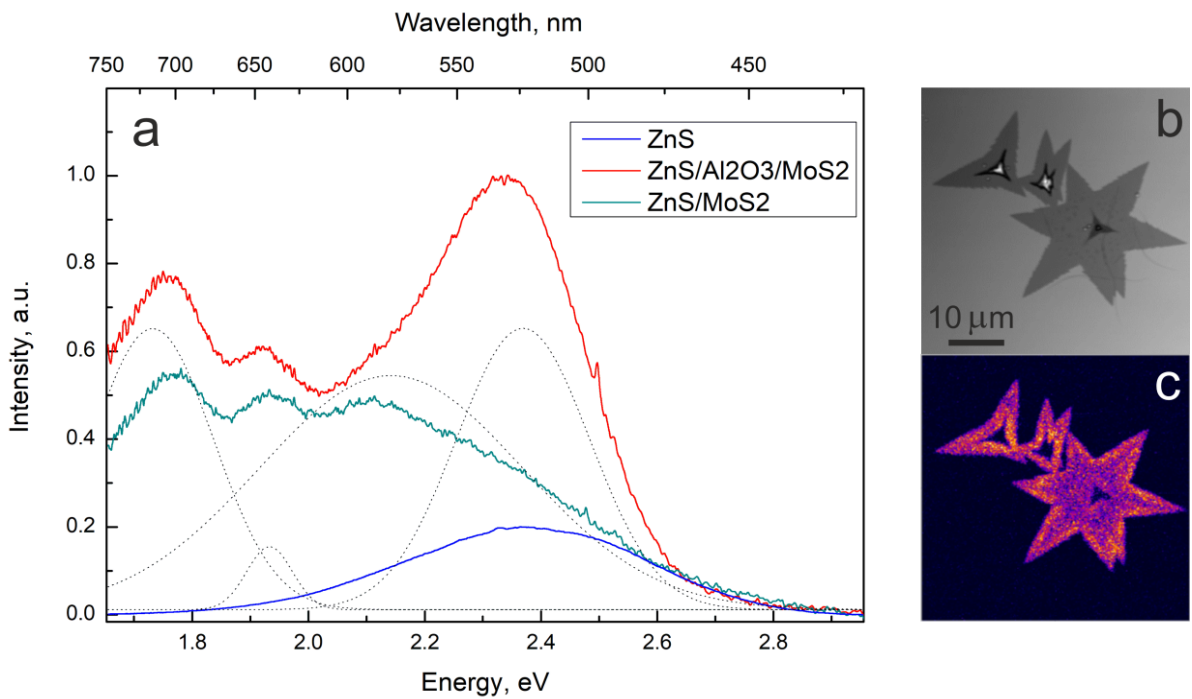

*Figure 5. Room-temperature photoluminescence (PL) spectra at 266 nm excitation wavelength for the as-synthesized ZnS NWs, as well as ZnS/MoS$_2$ and ZnS/Al$_2$O$_3$/MoS$_2$ (ZnO/MoO$_3$ and ZnO/Al$_2$O$_3$/MoO$_3$ annealed in the sulphur atmosphere at 750°C) NWs (a). The PL spectrum of ZnS/Al$_2$O$_3$/MoS$_2$ NWs as a function of energy was decomposed on the energy scale into a set of Gaussian functions (dotted lines) to show separate contributions. The PL intensity is depicted in arbitrary units and does not contain information about relative intensities between the measured spectra. Confocal microscope images of MoS$_2$ microcrystals on Si/SiO$_2$ wafer: reflection mode image (b) and photoluminescence image excited by 532 nm laser line and measured at 685 nm (c). Artificial colors in (b, c) encode signal intensity: bright color corresponds to high PL intensity.*

## 4. Conclusions.

In this study, the influence of the Al$_2$O$_3$ interlayer on the morphology of core-shell ZnS/Al$_2$O$_3$/MoS$_2$ NWs was explored. The synthesis of NWs involved the successive coating of ZnO NW templates by thin layers of Al$_2$O$_3$ and MoO$_3$ precursors, followed by annealing in a sulphur atmosphere at 750 ºC. The sulphurisation process resulted in the growth of the MoS$_2$ shell and a conversion of the ZnO NW core into the ZnS phase, which causes unwanted morphology perturbation of NWs in the absence of the Al$_2$O$_3$ interlayer. We found that the introduction of the Al$_2$O$_3$ interlayer prevents morphological changes in the ZnO core during conversion to ZnS, thus preserving the NW's straight shape. The role of the amorphous Al$_2$O$_3$ interlayer is to separate and prevent the interactions between the ZnO core and the MoO$_3$ shell during the sulphurization process. The formation of a thin crystalline MoS$_2$ shell was confirmed by XRD, XPS, TEM, Raman, and optical photoluminescence spectroscopies.

Our approach, based on the use of the Al$_2$O$_3$ protective interlayer, can be applied for the synthesis of other core-shell NWs with a TMD shell, where the core material (e.g. ZnO NWs) can be altered by the reactive chalcogenides (S or Se) at high temperatures.

**Acknowledgments**

Financial support was provided by Latvian Science Council grant Nr. lzp-2020/1-0261. E.B. and S.V. were supported by the European Union's Horizon 2020 program, under Grant Agreement No. 856705 (ERA Chair "MATTER"). The authors are grateful to Armands Ozols for assistance in the synthesis of nanomaterials. The Institute of Solid State Physics, University of Latvia (Latvia) as the Centre of Excellence has received funding from the European Union's Horizon 2020 Framework Programme H2020-WIDESPREAD01-2016-2017-Teaming Phase2 under grant agreement no. 739508, project CAMART2.